\documentclass[aps,prd,reprint,superscriptaddress,nofootinbib,floatfix,longbibliography]{revtex4-2}
\usepackage{amsmath}
\usepackage{amsfonts}
\usepackage{bm}
\usepackage{graphicx}
\usepackage{hyperref}
\usepackage{soul}
\usepackage{color}
\usepackage{comment}

\enlargethispage{\baselineskip}

\hypersetup{colorlinks   = true,
            urlcolor     = blue,
            citecolor    = blue,
            linkcolor    = blue,
            menucolor    = blue,
            anchorcolor  = blue,
            filecolor    = blue
}
\everymath{\displaystyle}

\begin{document}

\title{Neutron Stars Mass-Radius relations analysis in the Quintessence scenario}

\author{A. Campitelli}
\email{alfonso.campitelli95@gmail.com}
\affiliation{Dipartimento di Fisica ``E.R.\ Caianiello'', Universit\`a degli Studi di Salerno,\\ Via Giovanni Paolo II, 132 - 84084 Fisciano (SA), Italy}

\author{L. Mastrototaro}
\email{lmastrototaro@unisa.it}
\affiliation{Dipartimento di Fisica ``E.R.\ Caianiello'', Universit\`a degli Studi di Salerno,\\ Via Giovanni Paolo II, 132 - 84084 Fisciano (SA), Italy}
\affiliation{Istituto Nazionale di Fisica Nucleare - Gruppo Collegato di Salerno - Sezione di Napoli,\\ Via Giovanni Paolo II, 132 - 84084 Fisciano (SA), Italy}

\begin{abstract}
In this paper, we explore the effects of General Relativity modification on the Mass-Radius relations of Neutron Stars induced by the presence of the Quintessence field. We consider, in particular, the Kiselev model, according to which the Quintessence field, being present in the entire Universe, might also be present around massive objects. Considering the Equation of State (EoS) for Baryonic matter BSk22 derived by A. Y. Potekhin et al., we infer the upper limit for NS masses in the presence of Quintessence. The presence of Quintessence generates a peculiar effect for which the Mass-Radius relation is unvaried and therefore the presence of Quintessence is indistinguishable from ordinary matter, at least for the Kiselev model studied in this paper.
\end{abstract}

\date{\today}
\maketitle

\section{Introduction}
\label{sec:introduction}
Recent cosmological observations indicate that our Universe is presently undergoing accelerated expansion, as suggested by various studies \cite{riess,riess1,riess2,riess3,riess4,riess5}. To elucidate this phase in the evolution of the Universe, several alternative or modified theories of gravity have been proposed. These theories aim to address shortcomings in the cosmological framework based on General Relativity, which serves as the standard cosmological model. One approach involves the incorporation of higher-order curvature invariants, allowing for inflationary behaviour and providing solutions to the flatness and horizon problems \cite{starobinski,starobinski1}. For a broader range of applications and alternative models, refer to various references \cite{Capozziello:2011et,Tino:2020nla,cosmo1,cosmo2,cosmo3,cosmo4,cosmo5,cosmo6,cosmo7,cosmo8,cosmo9,cosmo10,cosmo11,cosmo12,cosmo13,cosmo14,cosmo15,odi,cosmo17,cosmo19,cosmo20,cosmo21,cosmo22,cosmo23,cosmo24,cosmo25,Nojiri:2017ncd,Benetti:2020hxp,Bernal:2020ywq}. These alternative theories and related concepts stem from the realization that in the high curvature regime, curvature invariants become essential for constructing self-consistent effective actions in curved spacetime \cite{birrell, shapiro, barth}.

Another avenue to explain the accelerated expansion involves invoking a mysterious component known as Dark Energy. Despite observational support, the nature and origin of Dark Energy remain the subject of lively debate. One potential candidate for Dark Energy is Quintessence \cite{Jamil:2014rsa}, characterized by the ability to generate negative pressure. Given its ubiquitous presence throughout the Universe, Quintessence has the potential to drive the observed accelerated phase. The presence of a Quintessence field diffused in the Universe raises the possibility that it might also be present around massive gravitational objects, causing deformation of the spacetime surrounding them. This possibility was explored by Kiselev \cite{Kiselev:2002dx}, where the author solved the Einstein field equations for static spherically symmetric Quintessence surrounding a Black Hole in four dimensions. The study also demonstrates that under appropriate conditions of the energy-momentum tensor, the known solutions corresponding to the electromagnetic static field and the cosmological constant can be recovered.\\
The Kiselev solution introduces two parameters: the parameter of state $w_q$, constrained by $-1<w_q<-1/3$, and the Quintessence parameter $c$. The study of quintessential Black Holes is also motivated by M-theory/Superstring-inspired models \cite{Belhaj:2020oun,HeydarFard:2007bb,HeydariFard:2007qs}. Various applications and implications can be found in references \cite{Toshmatov:2015npp,Abdujabbarov:2015pqp,Ghosh:2015ovj,Jamil:2014rsa,Belhaj:2020rdb,Khan5essence,Abbas:2019olp,Khan:2020ngg,Javed:2019jag,Khan:2020pgl,Uniyal:2014paa}.\\
This paper aims to use numerical calculations to obtain Mass-Radius relations of Neutron Stars (NS) in the Quintessence model and then confront this to find if could fit the observations. Indeed, NS is a perfect environment to test gravity modification. NS are the remnants of massive stars that have left the Main Sequence after burning all the hydrogen and helium, exploding in Supernovae. The typical NS has a radius of a few tens of km, where all the matter that before the explosion belonged to the core of the original star is concentrated. This particular characteristic causes the NS to have an intense gravitational field and every particle within it to be subjected to high pressure. The energies of this system are far beyond those achievable on a planet. \\
The work is organized in the following manner: in Sec.~\ref{TOV equations for quintessence background} we present the results of Tolman – Oppenheimer – Volkoff (TOV) equations applied to Quintessence theory and in Sec.~\ref{Conclusions} we describe our conclusions.

\section{TOV equations for quintessence background}
\label{TOV equations for quintessence background}

The most general expression of the spherically symmetric line element can be written in the form:
\begin{equation}
ds^2 =  e^{\nu (r)}dt^2 - e^{\lambda (r)}dr^2  - r^2d\theta^2 - r^2\sin^2\theta d\phi^2 \, \ .
\label{eq:metric}
\end{equation}

We point out that $\nu(r)$ and $\lambda(r)$ are generic functions which can be specified by the choice of the matter. In the case of Quintessence, it is necessary to consider the presence of a tangential component of the pressure, hence the energy-momentum tensor is written in the form:
\begin{equation}
T^{\mu \nu} = \mathrm{diag}(-\rho_T, P_r, P_t, P_t) \, \ ,
\label{eq:tensor_q}
\end{equation}

where $\rho_T=\rho+\rho_q$ is the total energy density, $\rho$ and $\rho_q$ are the Baryonic and Quintessence densities respectively and $P_r$ and $P_t$ are the radial and tangential components of the pressure respectively. We report that $P_r$ is the pressure component that is naturally derived from the Equation of State of the ordinary matter, while we would need an additional equation for $P_t$ to solve the system. \\
Considering the matter that constitutes the NS as a perfect fluid under the assumption that does not support transverse stresses and has no mass motion, the matter energy-momentum tensor is given by:
\begin{equation}
{T_0}^0 = \rho \, \ ,  \, \ {T_k}^0 = {T_0}^k = 0 \, \ ,  \, \ {T_i}^j = -P_r{\delta_i}^j \, .
\label{eq:T}
\end{equation}

Moreover, from the general form of the energy-momentum tensor in Eq.~\eqref{eq:tensor_q} it follows that the Quintessence has only the components ~\cite{Chen:2008ra,Lambiase:2020pkc}:
\begin{eqnarray}
T^{\;\;\;0}_{q\, 0}&=&T^{\;\;\;r}_{q \, r}=\rho_q,\label{t3} \\
T^{\;\;\;\theta}_{q\, \theta}&=&T^{\;\;\;\phi}_{q\, \phi}=-\frac{1}{2}\rho_q[3\omega_q+1]\label{t4}\,.
\end{eqnarray}

Using the metric and the energy-momentum tensor to solve Einstein field equations, a generalized TOV equation for the pressure is obtained \cite{Riazi_2016}:
\begin{eqnarray}
{dP_r(r) \over dr} = &-& {(P_r(r) + \rho(r)) (m (r)  + 4 \pi r^3 (P_r -\rho_q)) \over r (r - 2 m (r))} \nonumber \\ &+& {2 (P_r - P_t) \over r} \, \ ,
\label{TOV Gen}
\end{eqnarray}
where 
\begin{equation}
{d m(r) \over dr} = 4 \pi \left(\rho (P_r)+\rho_q(P_r)\right) r^2 \, \ .
\label{Eq:dm}
\end{equation}
Equation ~\eqref{TOV Gen} reduces to the standard TOV equation when $\rho_q = 0$ and $P_r = P_t$. Based on Ref.~\cite{Estevez-Delgado:2020oyl, Ghosh:2015ovj}, we assume the following equations for the tangential pressure and the energy density of the Quintessence:
\begin{gather}
P_t (r) = P_r (r) - {3 \over 2} (1 + \omega_q) \rho_q (r) \, \ , \\
\rho_q (r) ={3 \over 2} {\beta \omega_q \over ({r \over r_0})^{3 (1 + \omega_q)} + 1} \, \ .
\label{eq:rhoq}
\end{gather}
The equation of the energy density is presented as in Eq.~\eqref{eq:rhoq} to avoid divergences. The terms $\beta$ and $r_0$ are normalization constants, the first indicates the percentage fraction of Quintessence and the second is used to prevent divergent results in the numerical solution by making the denominator dimensionless.\footnote{In literature, the expression of Quintessence energy density is of the form: $\rho_q (r) ={3 \over 2} {\beta \omega_q \over r^{3 (1 + \omega_q)}}$. Such a form, however, provides numerical issues for the solutions of generalized TOV when $r$ is nearly zero. To overcome these problems, we modify the denominator by adding the constant $r_0^{3 (1 + \omega_q)}$; such a procedure allows us to eliminate the divergences for $r < r_0$. It must be pointed out that the modification does not alter substantially the results since the order of magnitude of $r$ at the boundaries of the star (relevant to our analysis) is bigger than $r_0$.} We are interested in obtaining the radial extension of the NS, namely $R$ and $M=m(R)$. $R$ is defined as the radius for which $P_r$ goes to zero. Finally, to solve the system of equations that we have obtained and to determine the mechanical equilibrium of the matter distribution inside of a NS, it is also necessary to impose an Equation of State (EoS) of the Baryonic component that makes up the star, for which we consider the Equation of State BSk22~\cite{Potekhin:2013qqa}:


\begin{widetext}
\begin{equation}\label{P BSk22}
\mathrm{log_{10}} P_r = K + {p_1 + p_2 \xi + p_3 \xi^3 \over 1 + p_4 \xi} \{\mathrm{exp}[p_5 (\xi - p_6)] + 1\}^{-1} 
+ (p_7 + p_8 \xi) \{\mathrm{exp}[p_9 (p_6 - \xi)] +1\}^{-1} 
\end{equation}
\begin{equation}
+ (p_{10} + p_{11} \xi) \{\mathrm{exp}[p_{12} (p_{13} - \xi)] + 1\}^{-1} 
+ (p_{14} + p_{15} \xi) \{\mathrm{exp}[p_{16} (p_{17} - \xi)] + 1\}^{-1} 
+ {p_{18} \over 1 + [p_{20} (\xi - p_{19})]^2} + {p_{21} \over 1 + [p_{23} (\xi - p_{22})]^2} \,,
\nonumber
\end{equation}
\end{widetext}

where $\xi \equiv \mathrm{log_{10}} (\rho/\mathrm{g} \, \ \mathrm{cm^{-3}})$, while the parameters $K$ and $p_i$ are tabulated.

One of the methods that can be used to solve these equations is to treat the system like a boundary problem, in which the external boundary of the matter distribution is the following: the value of $r = R$ for which $P_r = 0$, while for $r < R~ \mathrm{and}~ P_r > 0$, the solution depends on the EoS. We have chosen the  BSk22 EoS because it verifies the conditions of the biggest NS ever discovered and observed, PSR J0740+6620, which has a mass equal to $M = 2.076^{+0.067}_{-0.067} \, \ \mathrm{M_{\odot}}$ with an estimated radius of $R = 12.89^{+1.26}_{-0.97}~\mathrm{km}$ \cite{Riley_2021, Salmi:2022cgy}.

At this point, it is possible to numerically solve the generalized TOV equations by setting the various parameters. The following plots are obtained by fixing the normalization constants $r_0 \sim\mathcal{O} (10^{-18}~\mathrm{km})$ and $\beta$, which we decided to vary by $10 \,\%$ to verify the stability within short steps. To obtain every point on the curve, we started by fixing one precise value of $\rho$, then using the BSk22 EoS we extracted the initial pressure $P_r$ value and $\rho_q$ from Eq.~\eqref{eq:rhoq}.

\begin{figure}
\centering
\includegraphics[scale=0.33]{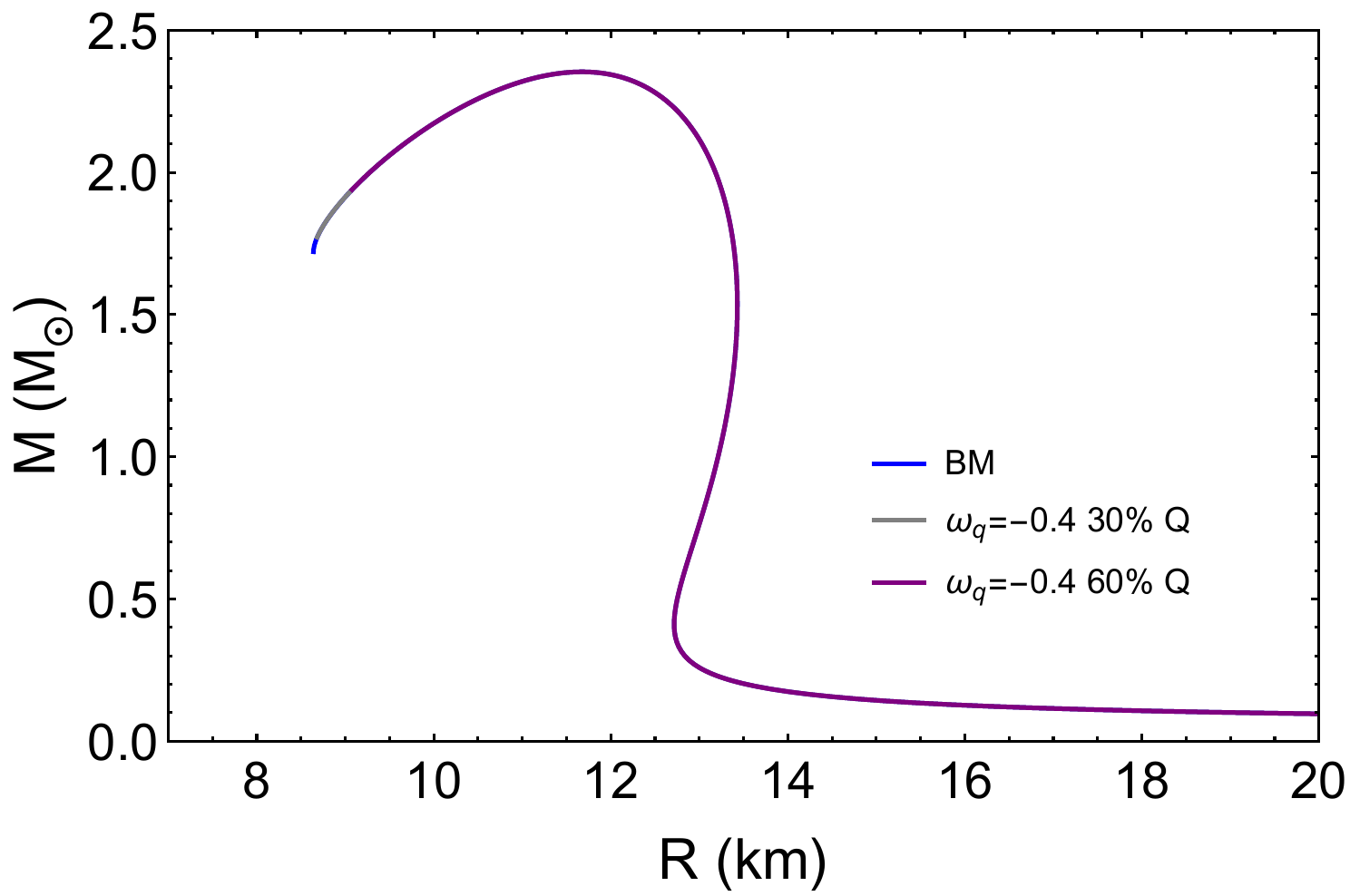}
\hspace{0.2cm}
\includegraphics[scale=0.33]{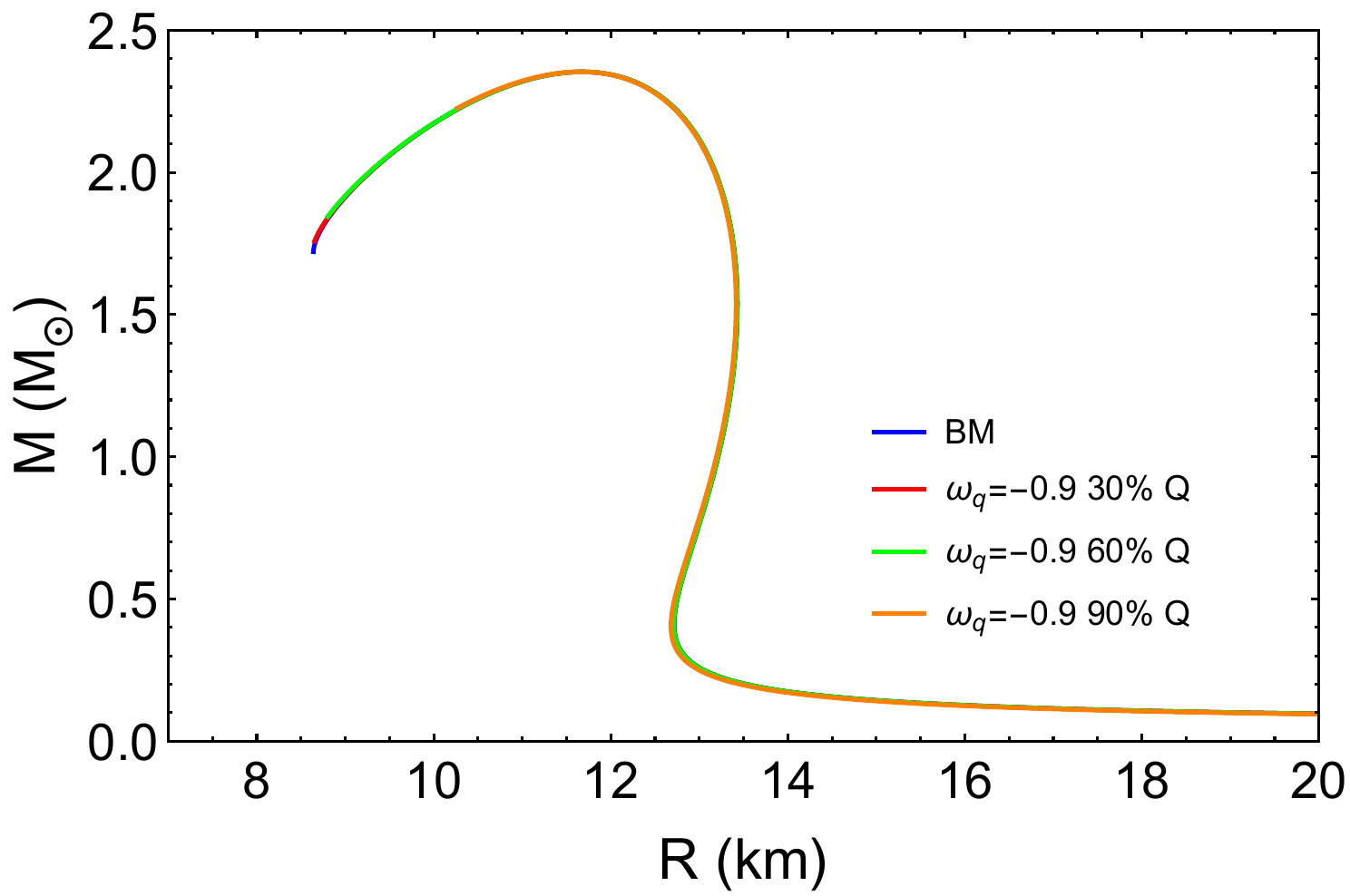}
\caption{Mass-radius stability curve Neutron Star in presence of Quintessence. Top: The Quintessence parameter is fixed to $\omega_q = - 0.4$, varying the percentage of Quintessence energy density with respect to BM up to two values, i.e. $30 \,\%$ and $60 \,\%$. Bottom: The same as Top with $\omega_q = - 0.9$, varying the Quintessence fraction up to three values, i.e. $30 \,\%$, $60 \,\%$ and $90 \,\%$.}
\label{fig1}
\end{figure}

\begin{figure}
\centering
\includegraphics[scale=0.3]{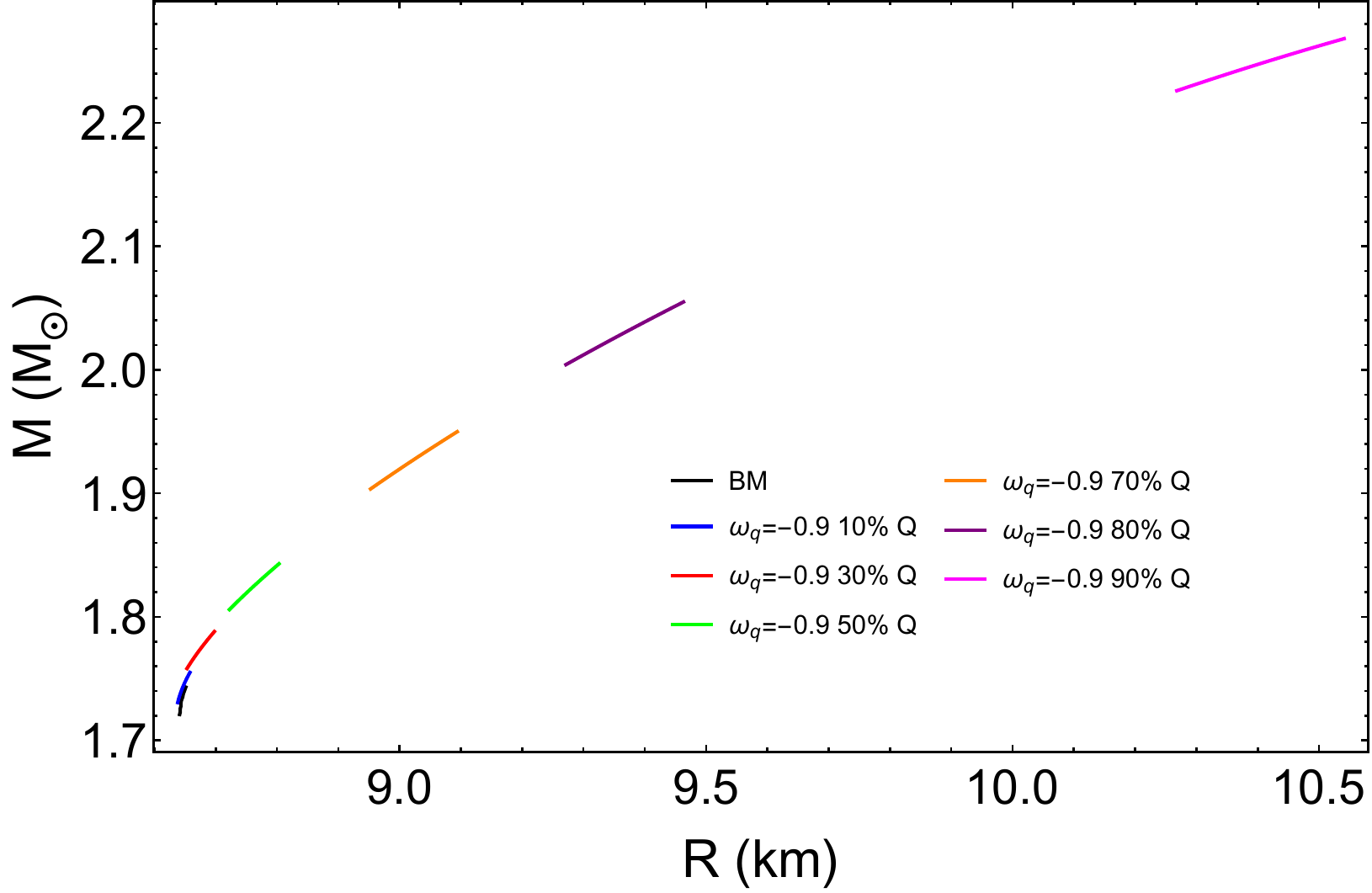}
\caption{Shift of the curve points for different percentages of Quintessence and fixed $\omega_q=-0.9$. The lines indicate the trend of the curve for ten values of BM energy density.}
\label{fig2}
\end{figure}

As shown in both the plots in Fig.~\ref{fig1}, the peculiar aspect that we want to stress is that the Quintessence component does not cause any significant variation in the Mass-Radius relation; in fact, it is shown that stars form almost the same way with a certain consistency, following the original profile curve of the Baryonic component. The only relevant aspect of the presence of the Quintessence is that the same initial conditions, i.e. pressure $P_r$ and energy densities, $\rho$ and $\rho_q$, do not lead to the same mass. This feature is more evident in the panel of Fig.~\ref{fig2}, which magnifies the discrepancy between the different percentages of Quintessence and values of $\omega_q$ as compared with the Baryonic Matter (BM). It appears clear that the more Quintessence is present inside the NS, the greater the shift will result from the curve of stability, following the same profile of the Baryonic case. The limit of this behaviour in the case of the figure is slightly higher than $90 \,\%$, after which instabilities appear and we do not obtain coherent values of $M$ and $R$.

It is possible to see that the plot corresponding to the value of $90 \,\%$ of Quintessence is not present for the case $\omega_q = - 0.4$ (Fig.~\ref{fig1}) because this quantity does not correspond to star formation. After $60\%$, the numerical calculation of the TOV equations gives rise to uncertainties, which means that the profile of the curve is no longer continuous. This happens because there are radii for which the total pressure is such that $(P_r -\rho_q)<0$, implying the absence of attraction.
This behaviour is caused by the fixed Quintessence profile defined by Eq.~\eqref{eq:rhoq}, which quickly decreases as $\sim r^{-3 (1 + \omega_q)}$. Therefore, the radial component of pressure $P_r$ is almost everywhere larger than $\rho_q$.

The fact that the NS consisting of both Baryonic matter and Quintessence is nearly equivalent to a NS made only by BM, means that it is impossible to deduce whether a star contains ordinary matter or a mixture with Dark Energy in the form of Quintessence.

\section{Conclusions}
\label{Conclusions}

In this paper, we have explored the Quintessence effect on Neutron Stars. We take into account the BSk22 EoS given in Ref.\cite{Potekhin:2013qqa} to test the upper limit for NS masses in the presence of Quintessence using a relatively stiff EoS. This EoS results in a Mass-Radius curve for GR with a maximum at $2.5 \sim M_{\odot}$ and radius of $R = 12.89^{+1.26}_{-0.97}~\mathrm{km}$.\\
Considering the presence of Quintessence, surprisingly we obtain a mimetic effect for which the Mass-Radius curve remains unchanged with respect to the GR profile. What changes is that, for the same $\rho(0)$, which corresponds to assuming the same initial conditions, we have different masses $M$ depending on $\rho_q(0)$, $\omega_q$, and, of course, on the percentage fraction $\beta$ of Quintessence. The new masses are shifted along the GR curve with respect to the Baryonic mass, labelled as BM in all Figures.\\
This effect appears relevant because it means that Quintessence does not alter the dimension of NS and is indistinguishable from the GR case as concerns the $M-R$ observations.

\vspace{2mm}
\acknowledgements
We warmly thank G. Lambiase for the discussions during the preparation of this work. We are grateful to L. Rezzolla  and Goethe University Frankfurt for 
hospitality during the preparation of part of this work.


\bibliography{sources.bib}

\end{document}